\documentclass{article}

\PassOptionsToPackage{numbers, compress}{natbib}

\usepackage[preprint]{neurips_2023}

\usepackage[utf8]{inputenc} 
\usepackage[T1]{fontenc}    
\usepackage{hyperref}       
\usepackage{url}            
\usepackage{booktabs}       
\usepackage{amsfonts}       
\usepackage{nicefrac}       
\usepackage{microtype}      
\usepackage{amsmath}
\usepackage{amsthm}
\usepackage{booktabs}
\usepackage{algorithm}
\usepackage{algorithmic}
\usepackage{amssymb}
\usepackage{graphicx,wrapfig}
\usepackage{booktabs}
\usepackage{multirow}
\usepackage[table,xcdraw]{xcolor}

\title{Stackelberg Decision Transformer for Asynchronous Action Coordination in Multi-Agent Systems}

\author{%
  Bin Zhang$^{1,2}$
  Hangyu Mao$^{3}$
  Lijuan Li$^{1,2}$
  Zhiwei Xu$^{1,2}$ 
  Dapeng Li$^{1,2}$ 
  Rui Zhao$^{3}$ \and
  \textbf{
  Guoliang Fan$^{1,2}$}\\
  $^1$Institute of Automation, Chinese Academy of Sciences\\
  $^2$School of Artificial Intelligence, University of Chinese Academy of Sciences\\
  $^3$ SenseTime Research\\
  \texttt{\{zhangbin2020, lijuan.li, xuzhiwei2019, lidapeng2020,}\\ \texttt{guoliang.fan\}@ia.ac.cn, \{maohangyu, zhaorui\}@sensetime.com}
}

\begin{document}

\maketitle

\begin{abstract}
Asynchronous action coordination presents a pervasive challenge in Multi-Agent Systems (MAS), which can be represented as a Stackelberg game (SG). However, the scalability of existing Multi-Agent Reinforcement Learning (MARL) methods based on SG is  severely constrained by network structures or environmental limitations. To address this issue, we propose the Stackelberg Decision Transformer (STEER), a heuristic approach that resolves the difficulties of hierarchical coordination among agents. STEER efficiently manages decision-making processes in both spatial and temporal contexts by incorporating the hierarchical decision structure of SG, the modeling capability of autoregressive sequence models, and the exploratory learning methodology of MARL. 
Our research contributes to the development of an effective and adaptable asynchronous action coordination method that can be widely applied to various task types and environmental configurations in MAS. 
Experimental results demonstrate that our method can converge to Stackelberg equilibrium solutions and outperforms other existing methods in complex scenarios.
\end{abstract}

\section{Introduction}

Multi-agent reinforcement learning (MARL) is a rapidly growing field with vast potential for practical applications, but it also poses significant challenges \cite{app_marl1, app_marl2}. 
In a multi-agent system (MAS), agents must not only maximize their individual rewards by interacting with the environment, but also dynamically coordinate with other agents to achieve  the optimal collective strategy \cite{lu2020algorithms}.  
This complexity leads to a series of issues that require resolution. 
From the perspective of task type, current mainstream methods for MARL primarily concentrate on fully cooperative tasks \cite{QMIX, HATRPO, MAT}, which are limited in their ability to handle complex interactions among agents and struggle with even simple coordination scenarios \cite{xu2023dual}.  
In mixed tasks, which are more generalized and widely applicable, there is both cooperation and competition among self-interested agents, each with private rewards, making it difficult to formally define and evaluate the quality of joint strategies. 
Environmental configuration is another issue, with most methods focusing on specific environment settings, such as agents acting based on local observations \cite{QMIX, MAT} or shared global states \cite{HATRPO, STEP, BiRL}. 
However, when the application scenario differs from the algorithms' given settings, results are often unsatisfactory. Therefore, it is imperative to explore new approaches that effectively address these issues and improve the versatility of MARL.

Notably, game theory offers an effective conceptual framework for resolving interactions among agents, making it a valuable tool for MARL \cite{NASH-Q, AQL, HATRPO}.
While significant progress has been made by incorporating the concept of equilibrium, the predominant focus has been on developing Nash equilibrium (NE) strategies, which assume that all agents make decisions simultaneously.
Nonetheless, this perspective neglects the asynchronous action coordination problem that commonly arises in realistic scenarios. 
To achieve optimal coordination strategies, 
it is essential for MAS to integrate social interactions or conventions (game structure) that incentivize agents to arrive at a corresponding equilibrium and establish a stable joint optimal policy.
This study focuses on the Stackelberg game (SG) model, which captures the hierarchical decision order among agents. Specifically, SG entails agents making decisions in a prescribed sequence, with leaders committing to their actions and followers discovering the optimal response to leaders' decisions. AQL \cite{AQL}, BiRL \cite{BiRL}, and STEP \cite{STEP} aim to learn Stackelberg equilibrium (SE) strategies. However, they typically impose stringent requirements on the network structure and environment, thereby constraining their scalability.

Recent advances in autoregressive sequence models \cite{TF} derived from natural language processing (NLP) \cite{GPT, Bert} have facilitated the development of novel reinforcement learning (RL) applications \cite{DT, TT}. 
These models have demonstrated remarkable performance and have been utilized to better comprehend and tackle RL issues. 
In this paper, we are delighted to discover that the hierarchical decision structure of SG is well-aligned with the form of the autoregressive sequence model. 
On the basis of this insight, we propose an approach that employs sequence models to address MARL issues. 

We begin by introducing a heuristic Stackelberg decision mechanism. 
Through the utilization of RL techniques and the construction of the decision form of SG, the convergence of the equilibrium strategy is achieved in a natural manner.
Consequently, we formally develop the Stackelberg Decision Transformer (STEER) based on spatio-temporal sequential Markov game (STMG) \cite{STEP} and Transformer in Transformer structure \cite{TIT,TITRL}. 
It processes distinct environment data via the Inner Transformer Block, while the Outer Transformer Block approximates the policy and value functions for each agent through autoregressive means. Our methodology offers a more logical and scholarly basis for investigating the dynamics of MARL, as well as a more comprehensive training paradigm. 
It can produce effective policies for fully-cooperative and mixed decision-making tasks. Furthermore, the comprehensive sequence modeling capability of the Transformer enables us to effectively manage tasks in a variety of environment configurations. Additionally, all agents update their networks simultaneously, reducing the computational costs and time constraints that are previously imposed by Stackelberg-based RL methods. We demonstrate the SE policy learning capability of STEER in single-step and multi-step matrix games and evaluate the algorithm using multi-agent MuJoCo \cite{ma-mujoco}, Google Research Football \cite{GRF}, and Highway On-Ramp Merging \cite{STEP} benchmarks. Our experimental results demonstrate that STEER outperforms potent baselines such as MAPPO \cite{MAPPO}, HAPPO \cite{HATRPO}, STEP \cite{STEP}, and MAT \cite{MAT} in terms of applicability and performance.

\section{Related Work}

\textbf{MARL. }
The most commonly used approaches in the MARL community focus on learning NE strategies under fully cooperative tasks, assuming that agents make decisions simultaneously. Value function decomposition methods \cite{QMIX}, for example, employ parameter sharing technique to train the same network for all agents. Other approaches consider learning heterogeneous strategies for agents, which better aligns with intuition, particularly in scenarios with self-interested or heterogeneous agents. HAPPO \cite{HATRPO} and A2PO \cite{A2PO} optimize the policy of each agent through a sequence update scheme.
Despite their advantages, these approaches have some drawbacks that must be addressed, including higher learning costs and extended training time. 
Although MAT \cite{MAT} can somewhat alleviate these issues, the advantage decomposition theorem on which they rely only applies to fully-cooperative scenarios, limiting their ability to handle diverse types of tasks. In situations where agents have private rewards, defining the joint advantage value becomes challenging, and evaluating the quality of the joint policy becomes difficult. 
In this paper, we aim to develop a universal approach for learning heterogeneous strategies in both fully-cooperative and diverse mixed scenarios.

\textbf{Stackelberg based MARL. }
Our research endeavors to address the prevalent challenge of asynchronous action coordination in MAS, with a particular emphasis on hierarchical coordination and SG structure among agents. 
Given the superiority of SE over NE in terms of existence, determinacy, and Pareto optimality \cite{SG, BiRL}, recent studies have delved into the application of SE in RL.
Similar to Nash Q-learning, AQL \cite{AQL} updates the Q-value function in an asymmetric setting by calculating the SE of the stage game at each iteration.
BiRL \cite{BiRL} proposes a two-player MARL method, utilizing a DQN-based \cite{DQN} learner for the leader and a DDPG-based \cite{MADDPG} learner for the follower. 
To enforce the SE policy, both the leader and follower need to store each other's model. 
STEP \cite{STEP} designs a complex structure to enable the execution of heterogeneous SE policies through parameter sharing, with followers inferring the actions of leaders to determine their response policies.
In order to comply with SG's requirements, these methods utilize intricate network structures and follow the presupposition that all agents share a common global state. This limitation narrows the scope of their applicability.
Our goal is to develop a more adaptable method that can overcome the stringent constraints imposed by the aforementioned methods on network structure and environment.

\textbf{Transformer in RL. }
Transformers \cite{TF} have acquired considerable traction in NLP, garnering the interest of researchers from other fields. Following the successful deployment of Transformers in computer vision \cite{vit, TIT}, researchers have increased their focus on  applying autoregressive sequence models to RL in various methods. 
Recent offline RL methods \cite{DT, TT} view RL as a decision sequence modeling problem in the time domain, thereby avoiding the problems posed by traditional RL methods that employ bootstrap Bellman errors. 
Instead, these methods use Transformers to predict actions based on information about preceding decision sequences, leading to sequential decision-making that is similar to time series forecasting \cite{lim2021time}. 
Transformer-based work has also gained attention in MARL. 
UPdet \cite{hu2021updet} concentrates on representation learning, with Transformer processing relationships between various entities in observations and matching them with subsets of the action space. 
MADT \cite{meng2021offline} employs Transformer to introduce the MAS field to the offline pre-training and online fine-tuning paradigm. 
MAT \cite{MAT} incorporates a standard Encoder-Decoder Transformer structure and employs the advantage decomposition theorem to solve fully cooperative tasks while anticipating convergence to NE policies. 
However, its design also restricts its capacity to manage varying environmental configurations and broader mixed tasks. 
Our method treats MARL as a spatio-temporal sequential decision problem, employing Transformers to model the decision-making order of agents on spatial decision sequences and RL methods for online training on temporal decision sequences.

\section{Preliminaries}

\subsection{Spatio-Temporal Sequential Markov Game}
\label{section:STMG}

The Spatio-Temporal Sequential Markov Game (STMG) \cite{STEP} is an evolutionary version of Markov Game (MG) based on SG. 
It is defined by the tuple $\Gamma\triangleq\langle\mathcal{I}, \mathcal{S}, \{\mathcal{A}^i\}_{i\in\mathcal{I}}, \mathcal{P}, \{r^i\}_{i\in\mathcal{I}}, \gamma, \{h^i\}_{i\in \mathcal{I}}\rangle$, 
where $\mathcal{I}$ represents the set of all agents with $|\mathcal{I}|=n$,
and $s \in \mathcal{S}$ represents the environmental state.
$a^i \in \mathcal{A}^i$ is the action of agent $i$ and the joint action space is $\mathcal{A}=\prod_{i=1}^{n}\mathcal{A}^i$.
$\mathcal{P}:\mathcal{S\times \mathcal{A}}\to \Omega(\mathcal{S})$ represents the state transition function of the environment, where $\Omega(X)$ denotes the set of probability distributions over $X$.
$r^i:\mathcal{S\times \mathcal{A}\to \mathbb{R}}$ is the reward function of agent $i$ and $\gamma$ is the discount factor.
$h^i$ denotes the decision priority of agent $i$ and $\mathcal{H}=\{h^1, ..., h^n\}$ is a prioritized permutation of agents.
At time step $t$, the agent with priority $h^i$ executes its strategy $\pi^{h^i}:\mathcal{S}\times \mathcal{A}^{h^1}\times \cdot\cdot\cdot \times \mathcal{A}^{h^{i-1}}\to\Omega(\mathcal{A}^{h^i})$ based on the subgame state $s_t^{h^i}=(s_t,a_t^{h^1},...,a_t^{h^{i-1}})$. The environment transitions to a new state $s_{t+1}\sim P(s_{t+1}\mid s_t,\boldsymbol{a_t})$ after receiving the joint action $\boldsymbol{a_t}=(a_t^1,...,a_t^n)$ and assigns private rewards $r^i(s_t,\boldsymbol{a_t})$ for each agent. The joint policy is represented by $\boldsymbol{\pi}\left({s}_{t}\right)=\prod_{i=1}^{n} \pi^{h^i}(s_t^{h^i})$. The transition function and the joint strategy determine the marginal distribution of the state at each time step, i.e., $s\sim \rho_{\boldsymbol{\pi}}$. Within this framework, each agent aims to maximize its own discounted cumulative reward $R^i(\tau)=\sum_{t=0}^T\gamma^tr^i(s_t,\boldsymbol{a_t})$ over a trajectory $\tau$ of length $T$. 
According to Bellman Equation, the state and action-state value function of agent $i$ in STMG can be written as:
\begin{equation}
\begin{aligned}
    Q^{h^i}_{\boldsymbol{\pi}}(s, a^{h^1:h^{i-1}},a^{h^i})&=\mathbb{E}_{s \sim \rho, \boldsymbol{a} \sim \boldsymbol{\pi}}[\sum\nolimits_{t=0}^{\infty} \gamma^{t} \cdot r^{h^i}_{t}(s_t,\boldsymbol{a_t})\mid \mathrm{s}_{0}=s,\boldsymbol{a}_{0}^{h^1:h^i}=\boldsymbol{a}^{h^1:h^i}],\\
    V^i_{\boldsymbol{\pi}}(s,a^{h^1:h^{i-1}})&=\sum\nolimits_{a^{h^i}\in\mathcal{A}^{h^i}}\pi^i(a^{h^i}|s,a^{h^1:h^{i-1}})Q^{h^i}_{\boldsymbol{\pi}}(s,  a^{h^1:h^{i-1}},a^{h^i}).
\end{aligned}
\label{equ:STMG_q}
\end{equation}
$A^{h^i}_\pi(s,a^{h^1:h^{i-1}},a^{h^i})=Q^{h^i}_{\boldsymbol{\pi}}(s, a^{h^1:h^{i-1}},a^{h^i})-V^i_{\boldsymbol{\pi}}(s,a^{h^1:h^{i-1}})$ represents the advantage function.
In certain environmental settings, agents may have access to localized observations $\{\mathcal{O}^i\}_{i\in\mathcal{I}}$ that are specific to each agent. Additionally, when $r^1=\cdot \cdot \cdot=r^n$, the task is considered a fully-cooperative task, otherwise, it is referred to as a mixed task. 

\subsection{Stackelberg Game}

The Stackelberg game (SG) \cite{SE} is a well-established game-theoretic framework that models hierarchical decision-making structures where some agents have advantages over others. 
Typically, such structures consist of leaders, who are superior agents capable of committing to their actions prior to other agents, and followers, who are inferior agents that must respond to the leaders' decisions. 
Leaders make decisions based on the assumption that followers will always react optimally to their actions. As an illustration, consider two agents whose leader and follower policies are denoted by $\boldsymbol{\pi}=(\pi^1, \pi^2)$. This can be formulated as a bi-level optimization problem:
\begin{equation}
\begin{aligned}
    &\max_{\pi^1 \in \Pi^1}\{\mathcal{J}^1(\pi^1,\pi^2)|\pi^2\in \arg\max_{\pi^{2'}\in\Pi^2}\mathcal{J}^2(\pi^1,\pi^{2'}) \}, \\
    &\max_{\pi^2 \in \Pi^2} \mathcal{J}^2(\pi^1,\pi^2), 
\end{aligned}
\label{equ:bilevel_op}
\end{equation}
where $\Pi$ represents policy space and $\mathcal{J}^i\left(\pi^1, \pi^2\right)=E_{s \sim \rho, \boldsymbol{a} \sim \boldsymbol{\pi}}\left[\sum_{t=0} \gamma^t r_t^i\left(s, a_t^1, a_t^2\right)\right]$ is the objective function of agent $i$. 
SE strategies, denoted as $\boldsymbol{\pi}^*=(\pi^{1^*},\pi^{2^*})$, correspond to the optimal solution of this bi-level optimization problem.
According to Section \ref{section:STMG}, we have:
\begin{equation}
\begin{aligned}
V^1_{\pi^{1^*},\pi^{2^*}}(s) &\geq V^1_{\pi^1,\pi^{2^*}}(s),\\
V^2_{\pi^1,\pi^{2^*}}(s,a^1) &\geq V^2_{\pi^1,\pi^2}(s,a^1).
\end{aligned}
\end{equation}

\subsection{Transformer}

Compared to classical neural networks, Transformer \cite{TF} has shown remarkable performance in modeling sequential data 
by leveraging a powerful \textit{self-attention} mechanism.
In formal terms, a \textit{self-attention} layer processes a sequential input of $d$-dimensional embedding vectors denoted by $\{\mathbf{x^i}\in \mathbb{R}^d\}_{i=1}^n$. For each input token $\mathbf{x}^i$, query vector $\mathbf{q}^i \in \mathbb{R}^{d_q}$, key vector $\mathbf{k}^i\in \mathbb{R}^{d_k}$, and value vector $\mathbf{v}^i \in \mathbb{R}^{d_v}$ are generated through linear mappings, where $d_q=d_k$.  The vectors of the entire sequence are then represented as matrices $\mathbf{X}\in \mathbb{R}^{n\times d}$, $\mathbf{Q}\in \mathbb{R}^{n\times d_q}$, $\mathbf{K}\in \mathbb{R}^{n\times d_k}$, and $\mathbf{V}\in \mathbb{R}^{n\times d_v}$, respectively. The output of the self-attention layer, denoted as $\mathbf{Y}\in \mathbb{R}^{n\times d_v}$, is obtained by computing a weighted sum of all values: $\mathbf{Y}=\text{softmax}({\mathbf{QK^{T}}}/{\sqrt{d_q}})\mathbf{V}$.

\section{Heuristic Stackelberg Decision Mechanism for MARL}
\label{4.1}

\begin{figure*}[t]
    \centering
    \includegraphics[width=4.7 in]{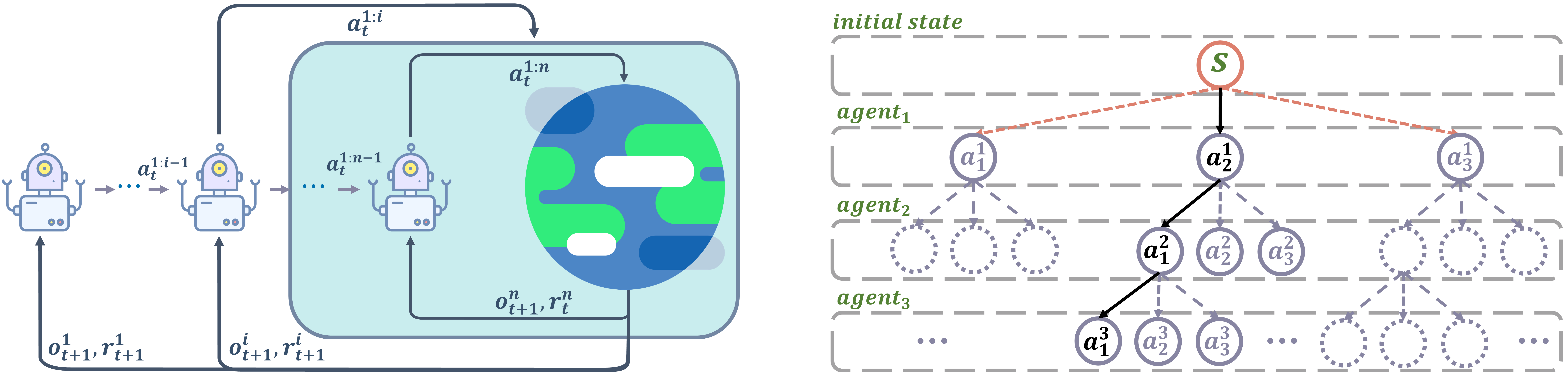}
    \vspace{-8pt}
    \caption{\emph{Left}: Exemplification of the heuristic Stackelberg decision mechanism for MARL. Followers interact with the environment based on joint actions with leaders, while leaders instruct followers as constituents of the environment.
    \emph{Right}: Schematic representation of spatio-temporal sequential decision-making. 
    }
    \label{fig:Mechanism}
    \vspace{-8pt}
\end{figure*}

In the context of multi-agent SG, we assign one agent to each priority level to match the spatio-temporal sequential decision pattern in STMG. This procedure bears resemblance to bi-level optimization in Equation~\ref{equ:bilevel_op} and yields an $n$-level optimization problem:
\begin{align}
    &\max_{\pi^i \in \Pi^i}\{\mathcal{J}^i(\pi^{1:i-1},\pi^i)|\pi^j\in \arg\max_{\pi^{j'}\in\Pi^j}\mathcal{J}^j(\pi^{1:j'-1},\pi^{j'}) \},  
    \label{equ:outer_op}\\
    &\max_{\pi^j \in \Pi^j} \mathcal{J}^j(\pi^{1:j-1},\pi^j),
    \label{equ:inner_op}
\end{align}
where $i\in[1:n]$ and $j\in[i+1, n]$.
Drawing inspiration from the heuristic algorithm for bi-level optimization \cite{bilevel-op, bilevel-op2}, we propose an RL-based heuristic Stackelberg decision mechanism (SDM) for this problem as shown in Figure~\ref{fig:Mechanism}. Within the hierarchical decision-making structure of STMG, each agent assumes the role of a follower to higher-level agents while simultaneously acting as a leader to lower-level agents. For followers, these inferior agents receive decision information from superior agents during both the execution and training procedures. The policy gradients of the agents are then updated in the direction of the optimal response to leaders, yielding an approximation of the solution to the inner optimization problem posed by Equation~\ref{equ:inner_op}. On the other hand, for leaders, these superior agents interact with the environment and perceive the reaction of the inferior agents. When updating their policies, leaders consider followers as part of the surrounding environment and maximize their own private rewards, resulting in an approximate solution to the outer optimization problem in Equation~\ref{equ:outer_op}.

Under the RL training paradigm, all agents possess the capability to maximize their individual utility in accordance with current conditions, thereby naturally achieving corresponding equilibrium. Through continuous interaction and trial-and-error with the environment, agents eventually arrive at a consensus wherein inferior agents execute optimal responses to the decisions of superior agents, and superior agents optimize their policies based on this premise, promoting the attainment of SE policies by all agents.

\begin{wrapfigure}{r}{0.37\textwidth}
\vspace{-10pt}
  \begin{center}
   \includegraphics[width=0.37\textwidth]{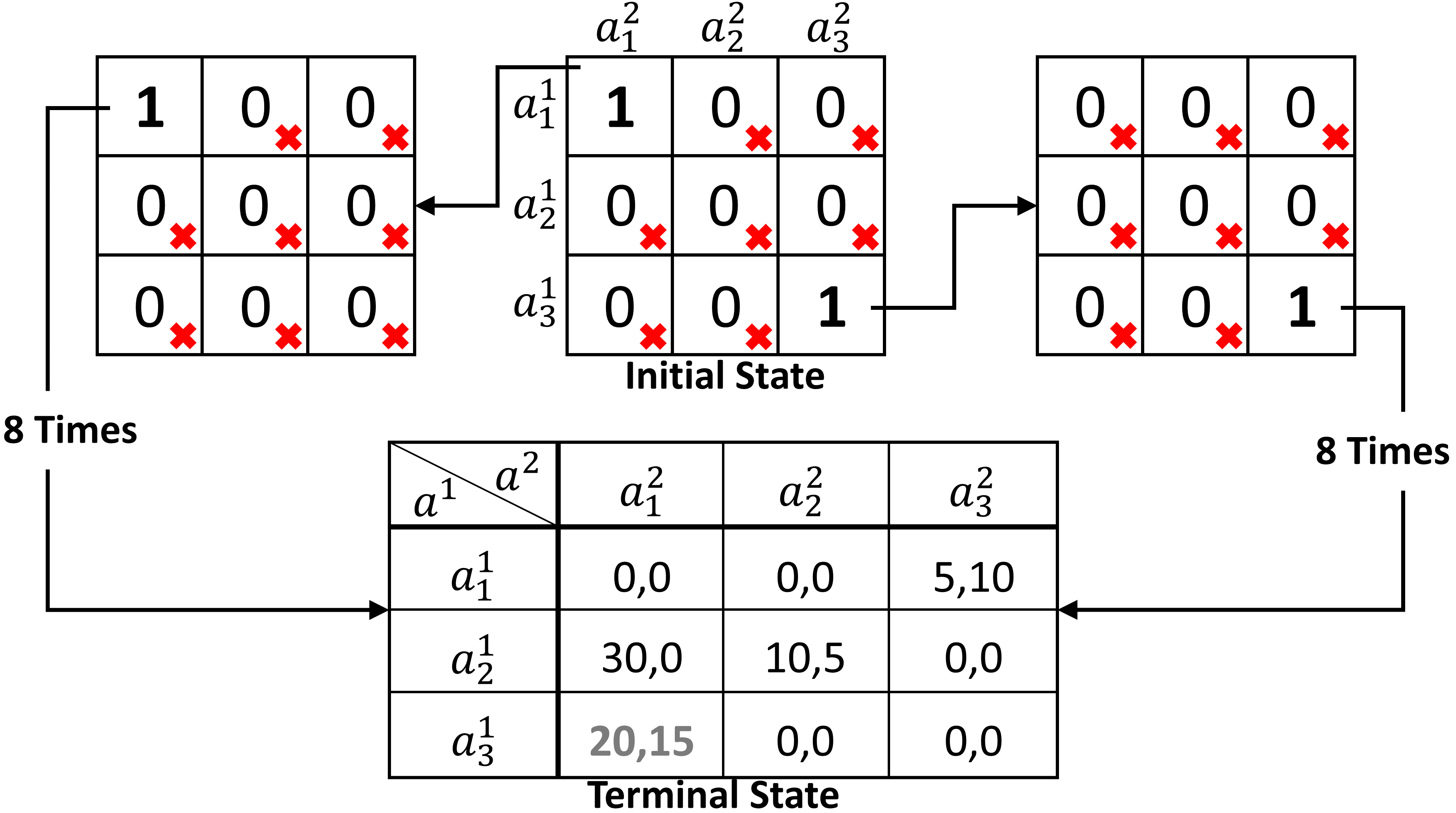}
  \end{center}
  \vspace{-10pt}
\caption{Multi-step matrix game: Coordination. 
Only actions with non-zero rewards are permissible before reaching the terminal state. } 
\vspace{-8pt}
\label{fig:msmg-coordination}
\end{wrapfigure}

Introducing SDM offers several advantages for resolving such coordination problems as opposed to assuming that agents act simultaneously.
For better illustration, we consider a simple two-agent three-action multi-step matrix game as shown in Figure \ref{fig:msmg-coordination}. 
To progress in the game, agents must make a choice between actions $(a^1_1,a^2_1)$ and $(a^1_3,a^2_3)$ at the initial state and repeat the same joint action until they coordinate at the endpoint to receive the final reward. Incorrect choices will result in game termination, and the agents will have to restart from the beginning.
Despite the game's simplicity, successful decision-making requires full cooperation and coordination between the agents to ensure the maximum total reward.
Our first point of emphasis is to clarify that the ideal space for the follower to act is narrowed down when the leader commits to its actions.
This constraint reduces the risk of both players pursuing disparate optimal strategies in the initial state, which may result in game failure.
As illustrated in Figure~\ref{fig:Mechanism}, this hierarchical decision-making process is similar to a depth-first search tree where the search space of child nodes is significantly simplified when parent nodes are fixed.
Secondly, from a game theory perspective, all three joint-actions $(a^1_1,a^2_3),(a^1_2,a^2_2),(a^1_3,a^2_1)$ are NE points in the final state, wherein neither player can enhance their payoff by changing their strategies. 
However, only point $(a^1_1,a^2_3)$ is the unique SE point, which results in the highest average payoff for both players. The detailed process for finding the SE can be found in the appendix.

This example enables us to conclude that in order to solve the SE strategy through MARL, certain requirements must be met.
All MARL methods require that first and foremost, all agents have an accurate perception of the current state. In addition, the direction of the joint policy's optimization is determined by the leaders' decision, and the duty of the followers is to learn the optimal response. As such, it is essential that agents have the ability to precisely evaluate the value function of the present sub-game state, which requires taking into account both the state of the environment and the leaders' decision-making information.

\section{Multi-Agent Stackelberg Decision Transformer}

\begin{figure*}[t]
    \centering
    \includegraphics[width=4.7 in]{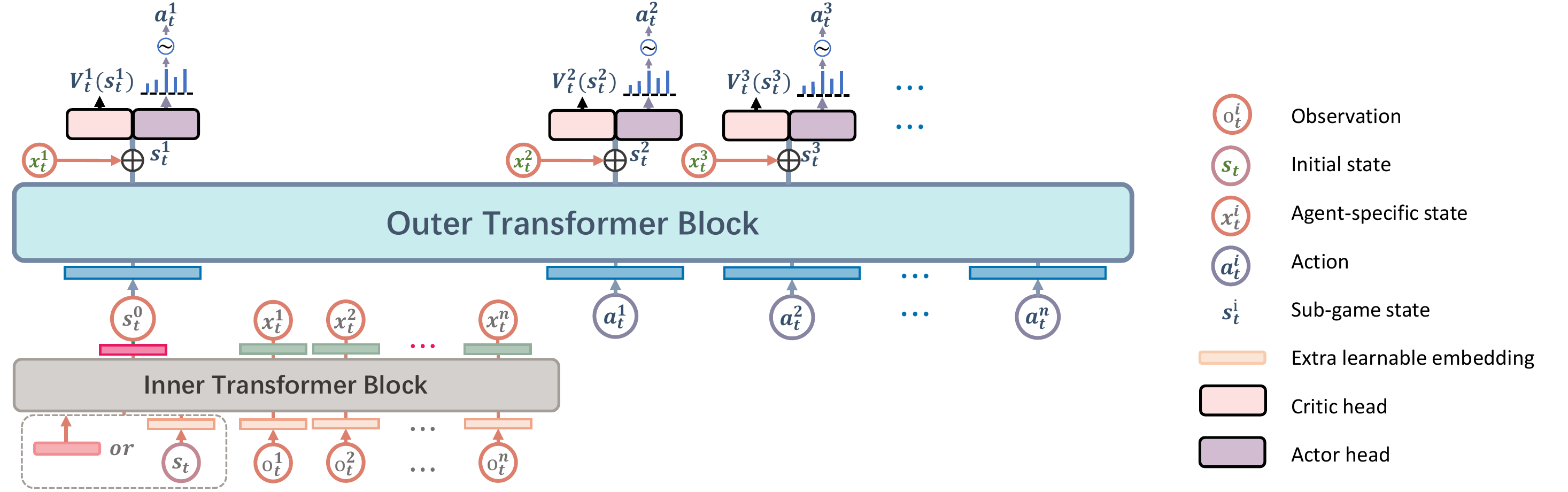}
    \caption{The overall architecture of STEER. At each time step, the state $s_t$ and observation $\{o_t^i\}_{i\in\mathcal{I}}$ information in different environmental configurations are transformed into agent-specific state embeddings $\{x_t^i\}_{i\in\mathcal{I}}$ through the Inner Transformer Block. Subsequently, each agent generates its action $\{a^i_t\}_{i\in\mathcal{I}}$ and sub-game state value function $\{V_t^i(s_t^i)\}_{i\in\mathcal{I}}$ in an autoregressive manner according to their priority level. The perception of the leaders' actions by agents is achieved through the Outer Transformer Block. }
    \label{fig:framework}
    \vspace{-10pt}
\end{figure*}

Integrated the discussions and studies made above,
we propose a formal solution called the Stackelberg Decision Transformer (STEER). 
In SDM, agents make decisions sequentially based on their priority, which corresponds precisely to the modeling structure of autoregressive sequential models.
To this end, we develop a hierarchical Transformer structure depicted in Figure~\ref{fig:framework} that takes inspiration from the Transformer in Transformer (TIT) architecture \cite{mao2022transformer}. 
Specifically, STEER employs the Inner and Outer Transformer Blocks, with both having sequence lengths of $n+1$.
The Inner Transformer Block (ITB) is responsible for processing state information for various environmental configurations, 
whereas the Outer Causal Transformer Block (OTB) manages decision element information for subsequent policy and value function fitting.  These two modules effectively fulfills the requirements outlined in Section~\ref{4.1}, respectively. 
Next, we provide a detailed description of STEER. 
For simplicity, it is assumed that the priorities of the agents are assigned based on their agent \textit{ID}.

\textbf{Inner Transformer Block. }
In ITB, agents' observation vectors $\{o^i_t\}_{i\in \mathcal{I}}$ are initially mapped to embeddings $\{e^i_t\}_{i\in \mathcal{I}}$ for further processing. If the environment includes additional global state information, the state embedding $e^0_t$ is utilized as the first token. Alternatively, if only local observation information is available, an extra learnable embedding $e^0_t$ is appended to the first token, similar to the class token in ViT \cite{vit}. Consequently, the input to ITB is represented as $\boldsymbol{e}_{l_0, t}=[e^0_t, e^1_t, ...,e^n_t]+\boldsymbol{E}_{pos}$, where $\boldsymbol{E}_{pos}$ represents the position embedding. Using multi-head self-attention (MHSA), multilayer perceptron (MLP) and layer normalization (LN), we can write the $j$-th block of ITB as:
\begin{align}
\boldsymbol{e}'_{\ell_j,t}=& \operatorname{MHSA}(\operatorname{LN}(\boldsymbol{e}_{\ell_{j-1},t})) + \boldsymbol{e}_{\ell_{j-1},t},\\
\boldsymbol{e}_{\ell_j,t}=& \operatorname{MLP}(\operatorname{LN}(\boldsymbol{e}'_{\ell_j,t}))+\boldsymbol{e}'_{\ell_j,t}.
\end{align}
Assuming a total of $L$ blocks, the output of ITB can be written as:
\begin{align}
\boldsymbol{Y}^{ITB}_t=\operatorname{MLP}(\boldsymbol{e}_{L,t})=[s^0_t, x^1_t,...,x^n_t],
\end{align}
where $s^0_t$ is the global game state embedding, which is encoded by the output of the first token at the last block.  $\{x^i_t\}_{i\in \mathcal{I}}$ represent agent-specific state embedding for all agents. 
ITB offers a flexible and adaptable methodology for handling various environmental state configurations.
It facilitates the production of precise abstract representations of game scenarios.

\textbf{Outer Transformer Block. }
In OTB,  $s^0_t$ serves as the abstract representation of the current game state, 
which, along with the actions $\{a^i_t\}_{i\in \mathcal{I}}$ taken by each prioritized level agent, forms the input sequence $\boldsymbol z_{0,t}=[s^0_t,a^1_t,...,a^n_t]$. 
The input of the first block in OTB can be expressed as $\boldsymbol{z}_{l_0,t}=\operatorname{MLP}(\boldsymbol{z}_{0,t})+\boldsymbol{E}_{pos}$. 
Subsequently, OTB utilizes masked multi-head self-attention (MMHSA) to generate decision element information in an autoregressive manner. 
Similar to ITB, this process is summarized as:
\begin{align}
\boldsymbol{z}'_{\ell_j,t}=& \operatorname{MMHSA}(\operatorname{LN}(\boldsymbol{z}_{\ell_{j-1},t})) + \boldsymbol{z}_{\ell_{j-1},t},\\
\boldsymbol{z}_{\ell_j,t}=& \operatorname{MLP}(\operatorname{LN}(\boldsymbol{z}'_{\ell_j,t}))+\boldsymbol{z}'_{\ell_j,t},\\
\boldsymbol{Y}^{OTB}_t = & \operatorname{MLP}(\boldsymbol{z}_{L,t}).
\end{align}
The primary objective of OTB is to aid agents in processing decision information from leaders.
Combining the current state information of each agent with decision information from higher level agents to create the current sub-game state embedding, represented as $\{s^i_t\}_{i\in\mathcal{I}}=(\boldsymbol{Y}^{ITB}_t+\boldsymbol{Y}^{OTB}_t)[0:n-1]$.
This embedding is subsequently transmitted to the Critic head (CH) and Actor head (AH) for the recursive approximation of the value and policy functions of agents:
\begin{align}
V^i_t(s^i_t)=V^i_t(s_t,a_t^{1:i-1})=\operatorname{CH}(s^i_t),\\
a^i_t\sim \pi^i_t(s^i_t)= \operatorname{AH}(s_t,a_t^{1:i-1}).
\end{align}

\textbf{Training Paradigm. }
Our approach is trained through end-to-end RL, using PPO \cite{PPO} as the underlying algorithm, which is one of the most potent and well-known algorithms in the community. 
Assuming that Transformer blocks, Actor head and Critic head are parameterized by $\omega, \theta, \phi$, respectively.
Correspondingly, the network needs to maximize the clipping objective function:
\begin{align}
\mathcal{L}(\theta, \omega) =
\mathbb{E}_{t,i}[
\min(r^i_{\theta, \omega} \hat A^i_\pi,
clip(r^i_{\theta, \omega}, 1\pm \epsilon)
 \hat A^i_\pi)
+
\eta S(\pi_{\theta, \omega}^i(s,a^{1:n-1}))
],
\label{pi_obj}
\end{align}
where $r^i_{\theta, \omega}=\frac{\pi^i_{\theta, \omega}(a^i|s,a^{1:i-1})}{\pi^i_{\theta_{old}, \omega_{old}}(a^i|s,a^{1:i-1})}$, $S(\cdot)$ refers to the Shannon entropy, 
and $\eta$ is its coefficient. 
$\epsilon$ represents the clipping ratio, and $\hat A^i_\pi$ serves as an estimation of the advantage value function.
Furthermore, STEER also requires the minimization of empirical Bellman TD-error:
\begin{align}
\mathcal{L}(\phi, \omega) =
\mathbb{E}_{t,i}[
\max ((V^i_{\phi,\omega}\left(s^i\right)-R^i)^2,(\operatorname{clip}(V^i_{\phi,\omega}(s^i), V^i_{\phi_{o l d},\omega_{old}}\left(s^i\right)\pm\varepsilon)-R^i)^2)
],
\label{v_obj}
\end{align}
where $\varepsilon$ serves as the clipping ratio and $R^i$ represents the cumulative return.

It is worth noting that the process of action generation in the execution phase differs from that in the training phase. Specifically, during the execution phase, actions are generated autoregressively. In contrast, during the training phase, the joint action sequence of the agents is captured and stored in the replay buffer. This allows for parallel calculation and updating, leading to significantly increased training speed compared to other heterogeneous policy learning techniques. Additionally, the TIT structure is utilized to estimate the value function and policy function of agents simultaneously, which aligns with the Multi-Task Learning (MTL) concept \cite{mtl}. Our method correspond to the most basic hard parameter sharing learning method in MTL. Furthermore, a further improvement direction would be to utilize more advanced MTL methods to simultaneously learn the actor and critic networks.

In comparison to MAT, which employs the standard encoder-decoder structure with Transformer, STEER is better adapted for MARL due to its adaptability to various environmental configurations and tasks. While MAT has demonstrated promising results in locally observable environments, it struggles in shared-state environments where the encoder's outputs may be similar and used as query values in the decoder, which has a significant negative impact on the self-attention mechanism. In contrast, STEER is adept at handling this issue. 
Furthermore, MAT disregards the decision information of the preceding agent and solely employs the encoder output to fit the state value function. While the actor network takes advantage of this information, the faulty guidance of the critic hinders it from attaining convergence to SE strategy like STEER. 

\section{Evaluation}
We present an evaluation and analysis of the proposed STEER on various testing benchmarks. Specifically, we assess STEER's ability to converge to SE solutions in both single-step and multi-step matrix game scenarios. Furthermore, we investigate the performance of STEER in complex cooperative scenarios, utilizing the widely adopted multi-agent MuJoCo (MA-MuJoCo) \cite{ma-mujoco} and Google Research Football (GRF) \cite{GRF} benchmarks. Additionally, we examine the effectiveness of STEER in a non-fully cooperative scenario, namely the Highway On-Ramp Merging (HORM) \cite{STEP} scenario. Furthermore, we experimentally verify the algorithm's generality and the reliability of its structure. We compare STEER to several advanced and comparable MARL methods, including MAPPO \cite{MAPPO}, which utilizes parameter sharing, HAPPO \cite{HATRPO}, which is specifically designed for heterogeneous policy learning, MAT \cite{MAT}, which is based on Transformer architecture, and STEP \cite{STEP}, which is based on SG.

\subsection{Finding SE Solutions}
\begin{figure*}[t]
    \centering
    \includegraphics[width=4.7 in]{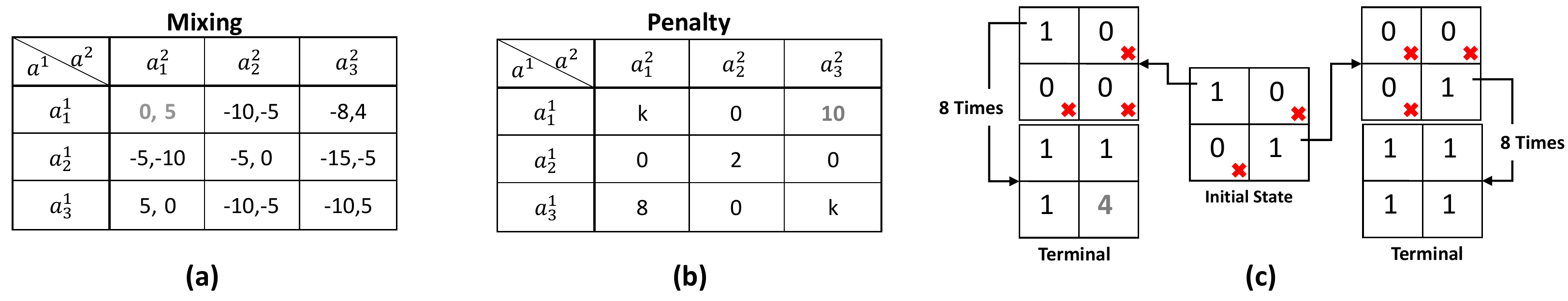}
    \vspace{-8pt}
    \caption{Matrix game scenarios. (a) Mixing. (b) Penalty ($k<=0$). (c) Multi-step matrix game: Cooperation. Multi-step matrix game: Coordination can be found in Figure~\ref{fig:msmg-coordination}.}
    \label{fig:toy_example_matrix}
    \vspace{-2pt}
\end{figure*}

\begin{figure*}[t]
    \centering
    \includegraphics[width=5.0 in]{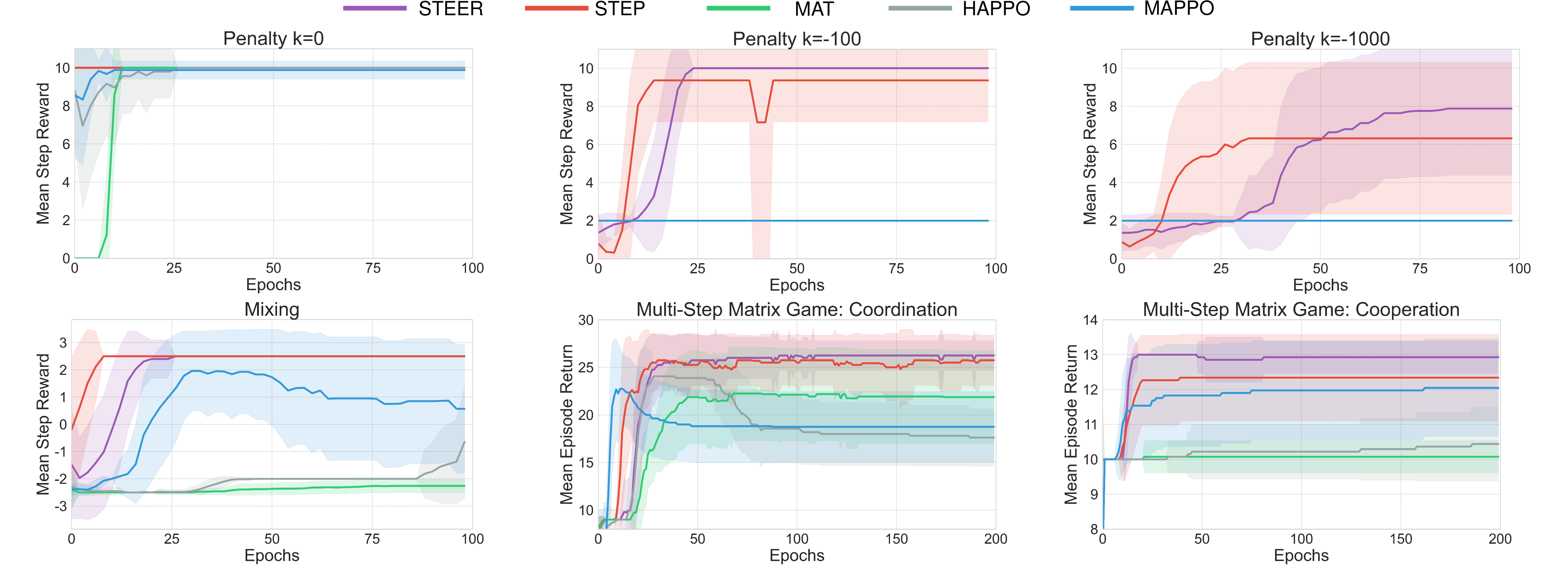}
    \vspace{-5pt}
    \caption{Performance comparison with baselines on matrix game tasks. A single standard deviation over trials is shaded.}
    \label{fig:toy_example_results}
    \vspace{-10pt}
\end{figure*}

\begin{table}[b]\scriptsize%
\centering
\vspace{-10pt}
\caption{
The percentage that converges to the global optimal strategy in Matrix game scenes.
}
\vspace{-5pt}
\begin{tabular}{@{}
>{\columncolor[HTML]{FFFFFF}}c 
>{\columncolor[HTML]{FFFFFF}}c 
>{\columncolor[HTML]{FFFFFF}}c 
>{\columncolor[HTML]{FFFFFF}}c 
>{\columncolor[HTML]{FFFFFF}}c 
>{\columncolor[HTML]{FFFFFF}}c 
>{\columncolor[HTML]{FFFFFF}}c @{}}
\toprule
{\color[HTML]{FFFFFF} } &
  \multicolumn{3}{c}{\cellcolor[HTML]{FFFFFF}Penalty} &
  \cellcolor[HTML]{FFFFFF} &
  \cellcolor[HTML]{FFFFFF} &
  \cellcolor[HTML]{FFFFFF} \\ \cmidrule(lr){2-4}
{\color[HTML]{FFFFFF} } &
  k=0 &
  k=-100 &
  k=-1000 &
  \multirow{-2}{*}{\cellcolor[HTML]{FFFFFF}Mixing} &
  \multirow{-2}{*}{\cellcolor[HTML]{FFFFFF}Coordination} &
  \multirow{-2}{*}{\cellcolor[HTML]{FFFFFF}Cooperation} \\ \midrule
{\color[HTML]{000000} STEER} & \textbf{100\%} & \textbf{100\%} & \textbf{72\%} & \textbf{100\%} & \textbf{95\%} & \textbf{96\%} \\
{\color[HTML]{000000} STEP}  & \textbf{100\%} & 93\%           & 44\%          & \textbf{100} & 94\%          & 90\%          \\
{\color[HTML]{000000} MAT}   & \textbf{100\%} & 0\%            & 0\%           & 0\%            & 46\%          & 5\%           \\
{\color[HTML]{000000} HAPPO} & \textbf{100\%} & 0\%            & 0\%           & 28\%           & 6\%           & 19\%          \\
{\color[HTML]{000000} MAPPO} & 95\%           & 0\%            & 0\%           & 63\%           & 14\%          & 65\%          \\ \bottomrule
\end{tabular}%
\label{tab:MixingResults}
\end{table}
To intuitively verify whether STEER converges accurately to the SE strategies, we conduct tests using single-step and multi-step matrix games, including fully-cooperative and mixed scenarios as illustrated in Figure~\ref{fig:toy_example_matrix}.
The results presented in Figure~\ref{fig:toy_example_results} and Table~\ref{tab:MixingResults} demonstrate that STEER outperforms other methods in all scenarios.
Notably, STEER consistently converges to SE solutions with the highest probability across all scenes. For instance, in the Penalty scenario, any deviation from the optimal strategy by an agent results in severe punishment for the other agent who has made the correct decision, rendering only the sub-optimal NE $(a^1_2,a^2_2)$ safe. Thus, as the penalty term k increases, it becomes increasingly difficult for agents to learn the optimal strategy. Nevertheless, within the STMG framework, the leader's decision information aids the follower in refining the ideal action space, resulting in the natural convergence to the optimal joint strategy by STEER. In contrast, all other algorithms, except for STEP and STEER, converge to the suboptimal solution with a 100\% probability when k < 0. Moreover, due to the capability of the Transformer, STEER remains effective even when k = -10000.
Notably, although MAT utilizes a similar sequential decision structure, it does not produce optimal results. Our analysis implies that this is a result of MAT's exclusive reliance on agents' local observation data rather than the sub-game state for approximating the value function, which can lead to erroneous guidance for actor updates. In addition, all methods except STEER and STEP are ineffectual when agents possess private rewards and must coordinate their actions.

\begin{figure*}[t]
    \centering
    \includegraphics[width=5.2 in]{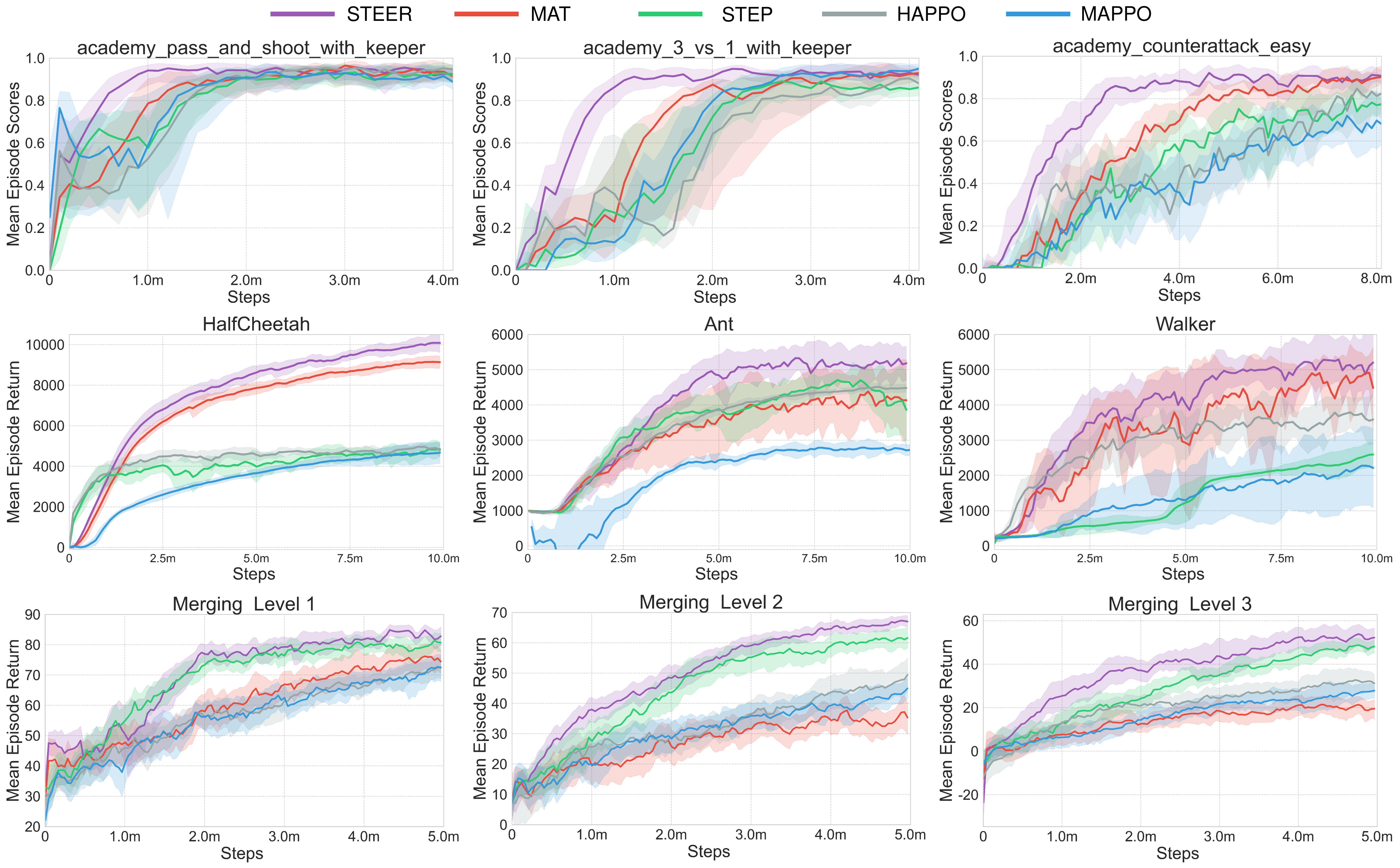}
    \vspace{-8pt}
    \caption{Comparing the evaluation performance on GRF, MA-MuJoCo and HORM. Error bars are a 95\% confidence interval across 5 runs.}
    \label{fig:complex_results}
\end{figure*}
\subsection{Performance in Complex Scenarios}

To evaluate STEER's efficacy in complex scenarios, we employ widely adopted benchmarks such as MA-MuJoCo, GRF, and HORM. 
These benchmarks include fully cooperative and mixed tasks, continuous and discrete control tasks, and tasks with different environmental state configurations. 
The appendix contains more thorough explanations of the environment.
STEER outperforms current state-of-the-art methods in all scenarios, as depicted in Figure~\ref{fig:complex_results}. These results highlight the superiority and adaptability of STEER in confronting complex scenarios.

The HORM scenarios are more intuitive for Stackelberg decision structure. 
One must first observe whether the vehicles on the main road are slowing down before deciding whether to merge into the lane. In MA-MuJoCo and GRF, all agents share the same reward. 
However, in HORM scenes, each agent wishes to pass through the intersection as quickly as possible while avoiding collisions, and they receive individual rewards from the environment.
In such coordinated scenarios, the joint advantage value function cannot be defined, and the advantage decomposition theorem is invalid, which directly leads to poor performance of MAT and HAPPO in HORM. In contrast, STEER perform better when dealing with coordination tasks.

\subsection{Ablation Studies}

\begin{figure*}[htbp]
    \begin{center}
    \includegraphics[width=5.4 in]{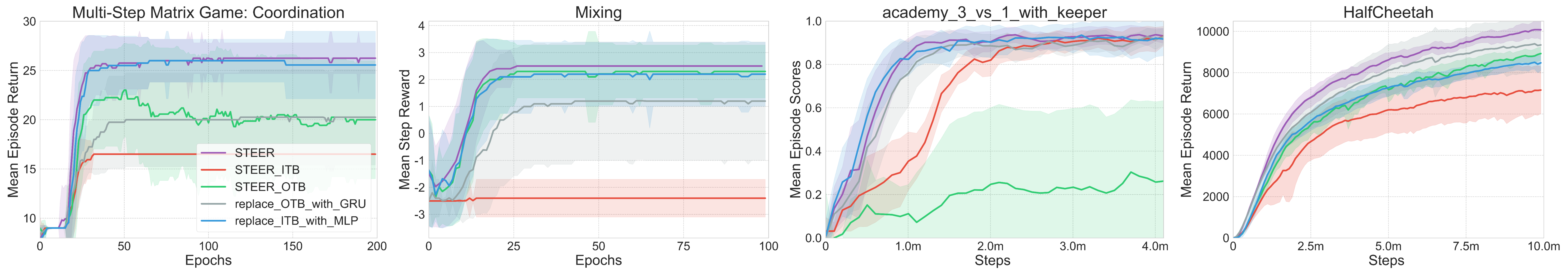}
    \vspace{-8pt}
    \caption{Performance comparison for different model architectures to explore the indispensability of each component.}
    \vspace{-20pt}
    \label{fig:ablation}
    \end{center}
\end{figure*}
STEER comprises two distinct modules, namely the ITB and OTB, responsible for processing state information and leaders' decision information, respectively, which together form an abstract representation of the current sub-game state. To ensure the reliability of the algorithm, we conducted ablation experiments on key components of STEER. This involves replacing the ITB with a simple MLP  (replace\_ITB\_with\_MLP), using a Recurrent Neural Network to replace the OTB (replace\_OTB\_with\_GRU) for decision information generation, and fitting the value and policy functions directly with the output from either the ITB (STEER\_ITB) or the OTB (STEER\_OTB). 
The experimental results, as shown in Figure~\ref{fig:ablation}, indicate that using only ITB for decision-making in the matrix game scenarios is equivalent to not introducing the SG structure, resulting in poor performance. Similarly, using only OTB for decision-making in complex scenarios is equivalent to agents lacking sufficient perception of the current environment and focusing more on the decision-making information of leaders, leading to algorithm failure. Moreover, replacing the Transformer with MLP or GRU also leads to performance degradation. Therefore, STEER satisfies all the requirements for model design outlined in Section \ref{4.1} and achieves optimal results.

\section{Conclusion and Future Work}
Our core insight is that the hierarchical decision-making structure of SG aligns perfectly with the modeling approach of autoregressive sequence models. Building upon this, we introduce the Stackelberg Decision Transformer method to solve the SE strategies of coordination tasks in MARL. Compared to previous work, our approach offers a more systematic and scholarly foundation for investigating the intricacies of MARL, as well as a more comprehensive training paradigm. Additionally, our method is more flexible in handling different environmental configurations, making it more applicable and scalable across different scenarios. As a fully centralized method, STEER can be easily extended to a decentralized execution system by using it as a teacher network and utilizing Knowledge distillation to train a separate policy network for each agent. We firmly believe that our method has broad potential for application in the MARL community. Furthermore, we suggest that exploring how to adaptively learn agent priority levels and how to use more advanced multi-task learning methods to simultaneously learn value and policy functions are worthy areas for further research.

\medskip

{
\small
\bibliography{neurips_2023}
\bibliographystyle{plainnat}

}

\newpage
\onecolumn
\appendix
\numberwithin{equation}{section}
\numberwithin{figure}{section}
\numberwithin{table}{section}
\renewcommand{\thesection}{{\Alph{section}}}
\renewcommand{\thesubsection}{\Alph{section}.\arabic{subsection}}
\renewcommand{\thesubsubsection}{\Roman{section}.\arabic{subsection}.\arabic{subsubsection}}
\setcounter{secnumdepth}{-1}
\setcounter{secnumdepth}{3}

\end{document}